\documentclass[%
 reprint,
 amsmath,amssymb,
 aps,prl
]{revtex4-1}

\usepackage{graphicx}
\usepackage{dcolumn}
\usepackage{bm}
\usepackage{hyperref}
\usepackage{color}
\usepackage{comment}
\usepackage{slashed}
\usepackage{simplewick}

\definecolor{deepblue}{rgb}{0.2,0.2,0.8}

\definecolor{deepred}{rgb}{0.8,0.2,0.2}

\newcommand{\hs}{\hspace{-0.2em}}

\newcommand{\vect}[1]{\boldsymbol{\mathbf{#1}}}

\def\deltabar{{\mathchar '26\mkern -10mu\delta}}

\newcommand{\ddbar}{{\rm d}\hspace*{-0.15em}\bar{}\hspace*{0.1em}}

\newcommand{\dd}{{\rm d}}

\begin{document}


\title{Stellar Basins of Gravitationally Bound Particles}

\author{Ken Van Tilburg}
 \email{kvt@kitp.ucsb.edu}
\affiliation{Kavli Institute for Theoretical Physics, University of California, Santa Barbara, California 93106, USA}

\date{\today}

\begin{abstract}
A new physical phenomenon is identified: volumetric stellar emission into gravitationally bound orbits of weakly coupled particles such as axions, moduli, hidden photons, and neutrinos.
While only a tiny fraction of the instantaneous luminosity of a star (the vast majority of the emission is into relativistic modes), the continual injection of these particles into a small part of phase space causes them to accumulate over astrophysically long time scales, forming what I call a ``stellar basin'', in analogy with the geologic kind. 
The energy density of the Solar basin will surpass that of the relativistic Solar flux at Earth's location after only a million years, for any sufficiently long-lived particle produced through an emission process whose matrix elements are unsuppressed at low momentum. 
This observation has immediate and striking consequences for direct detection experiments---including new limits on axion parameter space independent of dark matter assumptions---and may also increase the prospects for indirect detection of weakly interacting particles around compact stars.
\end{abstract}

\maketitle


\section{Introduction}
Stars are poor photon emitters. Their photon opacity is so high that the only effective radiating component is a thin shell near the stellar surface. Stellar energy losses per unit volume are thus suppressed by the surface-to-volume ratio. They are further diminished by thermal self-insulation: photon luminosity scales as the fourth power of the \emph{surface} temperature, which is typically several orders of magnitude lower than the \emph{core} temperature (a factor of about 2000 in the Sun).

Stars can thus serve as sensitive ``astrophysical laboratories'' of weakly coupled particles~\cite{raffelt1996stars,gamow1940possible, gamow1941neutrino}, whose contributions to the overall luminosity---via \emph{volumetric} emission---and thermal transport---via long mean free paths---can be disproportionately large. Indeed, neutrino emission is the main energy loss mechanism for the first $10^5\,\mathrm{y}$ after the birth of a neutron star~\cite{nomoto1987cooling}, despite neutrinos' tiny coupling. 
Even the Sun has a fractional luminosity of order $10^{-9}$ due to thermal neutrino-pair production~\cite{Vitagliano:2017odj}, in addition to the 3\% from fusion neutrinos.

Stellar cooling is also a powerful probe of weakly interacting, low-mass particles beyond the Standard Model of particle physics, for the same reasons. Such particles are motivated and predicted by wide classes of high-energy field theories as well as string theory~\cite{Arvanitaki:2009fg}, for example as moduli or pseudo-Nambu-Goldstone bosons of weakly broken global symmetries. The most notable of these is the QCD axion, in a framework that offers a dynamical explanation of the strong CP problem~\cite{Peccei:1977hh, axion1, axion2}. If these particles exist in the spectrum, they can---but need not---be the dominant component of dark matter (DM) in our Universe, with well-established early-universe production mechanisms~\cite{Preskill:1982cy, Abbott:1982af, Dine:1982ah,graham2016vector}. Leading constraints on the interactions of these exotic particles with ``regular'' matter often arise from the absence of anomalous cooling~\cite{raffelt1996stars,raffelt2008astrophysical,turner1990windows,an2015direct,hardy2017stellar} of the Sun~\cite{gondolo2009Solar,SCHLATTL1999353,vinyoles2015new}, horizontal-branch (HB) stars~\cite{ayala2014revisiting}, red giants (RG)~\cite{viaux2013neutrino}, white dwarfs (WD)~\cite{raffelt1986axion,blinnikov1994cooling,Bertolami:2014wua,Isern_2008,Isern_2009,isern2010axions,C_rsico_2012, C_rsico_2012b,C_rsico_2019}, neutron stars (NS)~\cite{ leinson2015superfluid,Hamaguchi_2018,
keller2013axions,Sedrakian_2016,Leinson_2014,Beznogov_2018}, and supernovae~\cite{Fischer_2016,Chang_2018,Carenza_2019,Page:2010aw,Shternin_2011}.

A vast number of experimental efforts (notable examples include~\cite{cast2017,Arisaka_2013,akerib2017first,fu2017limits,Wang_2020, Aralis_2020,aprile2019light,abe2018search,armengaud2018searches,aprile2017search}) are ongoing to detect (in a terrestrial laboratory setting) the relativistic Solar flux and a potential Galactic DM population of these particles. One leading experiment, sensitive to both types, recently reported~\cite{aprile2020observation} a statistically significant excess of electron-recoil events with energies just above a keV, intriguingly near the Sun's core temperature $T_\odot$. If the excess were in fact due to new fundamental physics beyond the Standard Model, many simple models appear in conflict with the aforementioned stellar cooling constraints, discounting these exotic hypotheses in favor of more mundane explanations such as unforeseen radioactive backgrounds.

In this work, I investigate a hitherto unknown effect in astroparticle physics that seems benign at first, but gives rise to dramatic observable consequences in laboratory experiments---including the limits and putative excess of Ref.~\cite{aprile2020observation}---and astrophysical observations. I stress that no new particle physics model is introduced; rather, I posit the occurrence of a new phenomenon that is generic within a large class of well-motivated models. The main point of this paper is simple: \emph{stars are able to emit massive particles into bound orbits.} The energy loss rate for this bound emission component is typically very small (see Eq.~\ref{eq:rhodotsimple}). However, this peculiar ensemble of particles populates parts of phase space that may survive for millions to billions of years around isolated stars, even in the inner Solar System. Over time, the density of this ``stellar basin'' can exceed the density of the relativistic stellar flux, including within the Solar System (cfr.~Eq.~\ref{eq:rhoparametric}).

In what follows, I describe how stellar basins of low-mass, weakly interacting particles can form and evolve. I discuss general aspects of soft emission near a particle mass threshold, before delving into a case study of Solar axions coupled to electrons. Based on these results, I present new limits on axion parameter space, and suggest the possible re-interpretation of the excess of Ref.~\cite{aprile2020observation} as due to a Solar axion basin. I conclude with potential avenues for future work.

\section{Stellar Basin Dynamics}
\subsection{Basin Formation}
\begin{figure}
\includegraphics[width = 0.5 \textwidth]{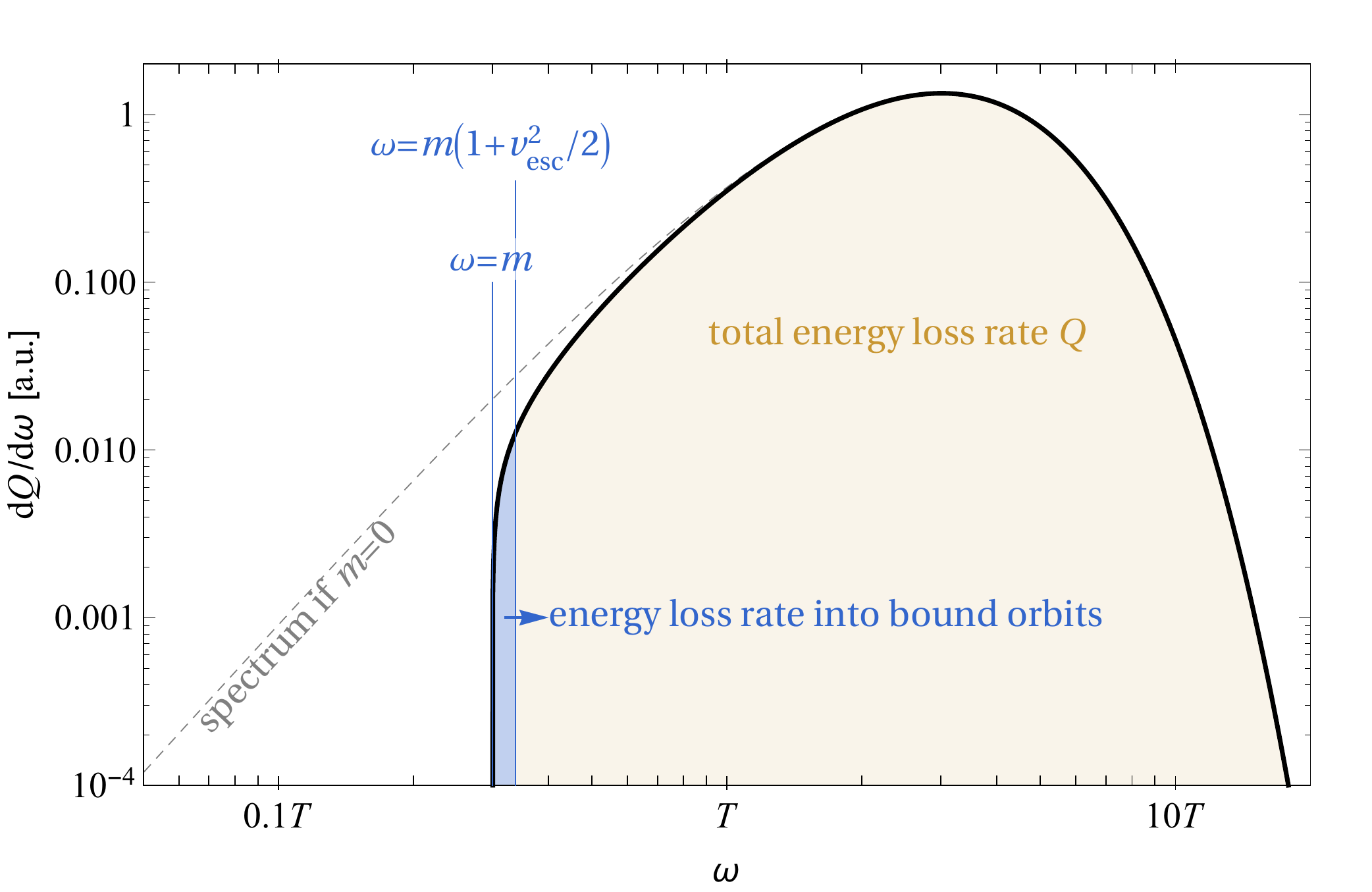}
\caption{A generic spectrum $\dd Q/\dd \omega$, the differential energy loss rate per unit volume per energy $\omega$ of the emitted particle, shown in black. If the particle has mass $m$, then stellar emission just above threshold, specifically $m < \omega < m(1+v_\mathrm{esc}^2/2)$ with $v_\mathrm{esc}$ the escape velocity, consists of nonrelativistic modes gravitationally bound to the star (blue region). While only a small fraction of the per-volume energy loss rate $Q$, given by the integral under the curve and most of which escapes to infinity (yellow region), the bound-orbit emission can accumulate over long times to form a ``stellar basin". The vertical axis has arbitrary units, and the escape velocity is greatly exaggerated for illustrative purposes. A massless spectrum is shown in dashed gray. The energy is in units of temperature $T$,
 with characteristic Boltzmann suppression at $\omega \gg T$.} \label{fig:spec}
\end{figure}

Consider a volume-emission process of a weakly interacting particle with mass $m$, with $\dd Q / \dd \omega$ the differential energy loss rate per unit volume per energy $\omega = \sqrt{m^2 + \vect{k}^2}$ of the emitted particle. The next two sections will explain how to calculate $\dd Q / \dd \omega$ in general and in a specific model, respectively; for now, assume it exists with some spectrum like the one shown in Fig.~\ref{fig:spec} for example. The total luminosity for this process is given by a volume integral over the stellar interior $L = \int \dd^3 R' \int_m^\infty \dd \omega \frac{\dd Q}{\dd \omega}$. For any physical process of interest, the vast majority of the luminosity is radiated away to infinity, leading to an energy density in unbound particles
\begin{align}
\rho_\infty(R) = \frac{L}{4\pi R^2} \label{eq:rhoinf}
\end{align}
that falls off as the inverse square radius $R$ outside the stellar surface at $R_*$.

However, a small fraction of the luminosity is emitted just above threshold ($\omega \simeq m$, $|\vect{k}| \ll m$), into bound orbits. Expanding in small kinetic energy per unit mass $\tilde{\omega}_k \equiv (\omega - m)/m \simeq \vect{k}^2/2m^2$, one finds in general that
\begin{align}
\frac{\dd Q}{\dd \tilde{\omega}_k} \simeq \sum_p \tilde{Q}_p(R') \tilde{\omega}_k^{n_p/2} + \dots  \label{eq:tildeQ}
\end{align}
where $n_p$ is a positive integer, the sum is over different emission processes $p$, and energy dependence is extracted out such that $\tilde{Q}_p$ only depends on the radius $R'$ within the stellar interior (assumed to be spherically symmetric).

Particles emitted into bound (radial) orbits are those with negative asymptotic energy per unit mass $\tilde{E} \equiv \tilde{\omega}_k + \Phi(R)$ where $\Phi(R)$ is the gravitational potential. The probability density $\mathcal{P}(R,\tilde{E})$ for a particle of normalized energy $\tilde{E}$ to be at radius $R$ is proportional to the inverse of its velocity $v(R,\tilde{E}) = \sqrt{2[\tilde{E}-\Phi(R)]}$:
\begin{align}
\mathcal{P}(R,\tilde{E}) = \frac{C(\tilde{E})}{v(R,\tilde{E})}; \qquad C(\tilde{E}) \simeq \frac{4 \sqrt{2}}{5 \pi^2} \frac{(-\tilde{E})^{7/2}}{(G_N M_*)^3}. \label{eq:pRE}
\end{align} 
The normalization constant $C(\tilde{E})$ is fixed by requiring $\int \dd^3 R \, \mathcal{P}(R,\tilde{E}) = 1$; the quoted expression is valid for a gravitational potential $\Phi(R) \simeq - G_N M_* / R$, and is a good approximation for orbits with maximum radius far outside the radius $R_*$ of the gravitating body.

The bound energy density at radius $R \ge R_*$ grows at a rate:
\begin{align}
\hspace{0em} \dot{\rho}_\mathrm{b}(R) = \sum_p \int \hspace{-0.2em}\dd^3 R' \hs \int_{\Phi(R)}^0 \hspace{-0.5em} \dd \tilde{E} \, \tilde{Q}_p(R') \frac{C(\tilde{E})}{\sqrt{2}} \frac{[\tilde{E} - \Phi(R')]^{\frac{n_p}{2}}}{[\tilde{E} - \Phi(R)]^{\frac{1}{2}}}, \label{eq:rhodot}
\end{align}
For $R \gg R_* = \max\lbrace R' \rbrace$, one can approximate $\tilde{E} - \Phi(R') \simeq -\Phi(R')$, and  $C(\tilde{E})$ as in Eq.~\ref{eq:pRE}, yielding the simplification:
\begin{align}
\dot{\rho}_\mathrm{b}(R) = \frac{7}{32 \pi}  \frac{G_N M_*}{R^4} \hs \int \hs \dd^3 R' \, \sum_p \tilde{Q}_p(R') \left|\Phi(R')\right|^{\frac{n_p}{2}}. \label{eq:rhodotsimple}
\end{align}
In most cases of interest, the integral above evaluates to a result of order $\sum_p L_p \min\lbrace m^3/T(0)^3, 1\rbrace |\Phi(0)|^{n_p/2}$ with $T(0)$ and $\Phi(0)$ the core temperature and gravitational potential of the star, and $L_p$ the luminosity for each process ($L\equiv \sum_p L_p$).

The crux of stellar basins is that the weakly coupled bound particles can accumulate for astrophysically long times, compensating for their lower production rate relative to the unbound flux from Eq.~\ref{eq:rhoinf}. Denote by $\tau$ the effective $1/e$ lifetime of a bound orbit at radius $R$. This lifetime is set by the most efficient of the three dominant processes for depleting the basin, namely (gravitational) ejection with lifetime $\tau_\mathrm{eject}$,  (effective) absorption with lifetime $\tau_\mathrm{abs}$, and radiative decay with lifetime $\tau_\mathrm{rad}$. More precisely, one has that $\tau^{-1} = \tau_\mathrm{eject}^{-1} + \tau_\mathrm{abs}^{-1} + \tau_\mathrm{rad}^{-1}$. The ejection lifetime in the Solar System at $R = 1\,\mathrm{AU}$ is at least $\tau_\mathrm{eject}(\mathrm{AU}) \gtrsim 10^7\,\mathrm{y}$ for example (see next section), often much faster than decay or absorption channels, as discussed later around Eqs.~\ref{eq:tauabsorb} and \ref{eq:taurad}.

After some time $> \tau$, the basin will be filled and reach a (quasi-)steady-state density $\rho_\mathrm{b} = \dot{\rho}_b \tau$. At this saturation point, the ratio of bound-to-unbound energy densities is:
\begin{align}
\frac{\rho_\mathrm{b}}{\rho_\infty} &= \frac{7}{8}  \frac{\tau}{R} |\Phi(R)| \frac{\int \dd^3 R' \, \sum_p \tilde{Q}_p(R') \left|\Phi(R')\right|^{n_p/2}}{L} \label{eq:rhoratio1}\\
&\sim  \frac{\tau}{R} v_\mathrm{esc}^2(R) v^{n_p}_\mathrm{esc}(0) \min\left\lbrace \frac{m^3}{T(0)^3} , 1 \right\rbrace. \label{eq:rhoratio2}
\end{align}
The second line contains the parametric estimate for the integral mentioned below Eq.~\ref{eq:rhodotsimple}, took only one dominant emission process with exponent $n_p$, and is written in terms of the escape velocity $v_\mathrm{esc}(R) \equiv \sqrt{-2 \Phi(R)}$. 

The salient feature of stellar basins---the large accumulation time $\tau$---is responsible for the huge enhancement in the first factor in Eq.~\ref{eq:rhoratio2}, which can easily outweigh the remaining small factors due to phase-space suppression. Near the surface of an isolated neutron star ($R_* \sim 10\,\mathrm{km}$), this factor can in principle be as large as its age over its light-crossing time, $\tau/R \sim 10^{21}$ at $\tau \sim 10^9\,\mathrm{y}$. Compact remnants like neutron stars or white dwarfs furthermore have large escape velocities, mitigating the phase-space suppression.

Even at Earth's location in the Solar System, the bound energy density in the basin may exceed that of the unbound Solar flux, for processes that are not suppressed near threshold (i.e.~$n_p = 1$) and for masses $m$ not far below the core temperature $T(0)$. Numerically, the escape velocities are $v^\odot_\mathrm{esc}(\mathrm{AU}) \approx 1.40 \times 10^{-4}$ and $v^\odot_\mathrm{esc}(0) \approx 4.61 \times 10^{-3}$, while $\tau/\mathrm{AU} \approx 6.3 \times 10^{11} (\tau / 10^7\,\mathrm{y})$. Eq.~\ref{eq:rhoratio2} therefore indicates that the bound population can start dominating the unbound Solar flux of particles with mass $m\sim T_\odot(0) \sim \mathrm{keV}$ after only a million years:
\begin{align}
\frac{\rho_\mathrm{b}}{\rho_\infty} \sim \frac{\tau}{10^6 \, \mathrm{y}} ~\left[\text{for $R = 1\,\mathrm{AU},~n_p = 1, ~m\sim T_\odot(0) $}\right] \label{eq:rhoparametric}
\end{align}
up to $\mathcal{O}(1)$ prefactors. Finally, the relative enhancement of the bound state population at Earth's surface from particle production in the core of Earth itself is even more striking. With surface and core escape velocities of $v_\mathrm{esc}^\oplus(R_\oplus) \approx 3.73 \times 10^{-5}$ and $v_\mathrm{esc}^\oplus(0) \approx 5.02 \times 10^{-5}$, one finds that $\rho_\mathrm{b}/\rho_\infty \sim 10^5 (\tau/10^9\,\mathrm{y})$ if $n_p= 1$ and $m \sim T_\oplus(0) \sim 0.5\,\mathrm{eV}$. In absolute terms, $\rho_\mathrm{b}$ in Earth's basin will turn out to be small, though it can equal or exceed that of the Solar basin at small masses $m \lesssim T_\oplus(0)$, depending on the ratio of gravitational ejection timescales.

\subsection{Gravitational Ejection}
A critical parameter determining the maximum energy density to which the stellar basin can fill up is the lifetime $\tau(R)$. Contributions to the inverse lifetime from the radiative decay rate $\tau_\mathrm{rad}^{-1}$ (Eq.~\ref{eq:taurad}) and from the effective absorption rate $\tau_\mathrm{abs}^{-1}$ (Eq.~\ref{eq:tauabsorb}) in the stellar interior are relatively straightforward to calculate, given a specific particle physics model. However, in an important case of interest, namely that of ultra-weakly coupled particles in the Solar System, the gravitational ejection timescale $\tau_\mathrm{eject}(R)$ is the principal limiting factor.

The main obstacle towards an accurate determination of $\tau_\mathrm{eject}(R)$ is due to the well-known chaotic nature of orbits in the Solar System, with a Lyapunov time of approximately 5~million years at $R = 1\,\mathrm{AU}$~\cite{batygin2008dynamical}. Taking this to be the timescale at which the entire phase space volume is explored efficiently, a conservative estimate would be that orbits migrate to an ejection-inducing orbital resonance with a characteristic timescale of two Lyapunov times:
\begin{align}
\tau_\mathrm{eject}^\odot(\mathrm{AU}) \sim 10^7\,\mathrm{y} \quad \text{(conservative)}. \label{eq:taueject1}
\end{align}
Even with such an overly efficient ejection rate, the parametric estimate of Eq.~\ref{eq:rhoparametric} predicts a local fractional overdensity of the stellar basin over the relativistic flux density by up to one order of magnitude.

At first glance, the estimate of Eq.~\ref{eq:taueject1} appears to be confirmed by numerical simulations of forward evolution of Near-Earth Objects (NEOs)~\cite{farinella1994asteroids,gladman1997dynamical}, with Ref.~\cite{gladman2000near} finding a median lifetime of 10~million years for 117 NEOs with perihelia $q < 1.3\,\mathrm{AU}$ and high-quality initial orbital elements. However, the vast majority of NEOs ended their lives due to Sun-grazing orbits (56\%) or planetary impacts (18\%), both of which leave weakly coupled particles unharmed; 16\% of NEOs survived the total integration time of 60~million years, while only 10\% were ejected from the Solar System~\cite{gladman2000near}. A fiducial $1/e$ ejection timescale would be
\begin{align}
\tau_\mathrm{eject}^\odot(\mathrm{AU}) \sim 10^8\,\mathrm{y} \quad \text{(fiducial)}, \label{eq:taueject2}
\end{align}
based on the NEOs unaffected by impacts, a restriction that could admittedly introduce bias.

There are four reasons that suggest even the fiducial value of Eq.~\ref{eq:taueject2} is an underestimate. First, the stellar basin is populated with vastly different orbits than those of asteroids or captured comet remnants. The main contribution to $\dot{\rho}_\mathrm{b}(\mathrm{AU})$ comes from directionally isotropic, radial orbits with aphelia very near $1\,\mathrm{AU}$ that are later processed---by secular perturbations from the planets---into high-eccentricity $e \approx 1$ elliptical orbits with semi-major axes $a \approx 0.5\,\mathrm{AU}$ at all inclinations $i$. In contrast, NEOs are preferentially located near the ecliptic plane $i \approx 0$, and near unstable parts of phase space, since they were resonantly driven away from the asteroid belt or (to a lesser extent) captured by close encounters in the inner Solar System~\cite{bottke2002debiased}. Second, the known NEO population is known to exhibit strong observational biases against (separately) high-$e$, high-$i$, and low-$a$ orbits, all of which are associated with higher dynamical lifetimes~\cite{bottke2002debiased,gladman2000near, rabinowitz1994population}. Third, the mortality rate of NEOs does not follow a simple exponential decay~\cite{gladman2000near}, indicating that various parts of phase space may be protected (e.g.~be locked into metastable Lidov-Kozai resonances~\cite{michel1996kozai}) and have longer ejection timescales. Fourth, since the basin energy density injection rate scales as $\dot{\rho}_\mathrm{b} \propto 1/R^4$, the total energy density in logarithmic radial shells (e.g.~between $R$ and $2R$) scales as $1/R$. So even if there were a depletion of $\rho_\mathrm{b}(R)$, orbits \emph{interior} to $R$ may be responsible for compensating replenishment as they migrate outward.

Indeed, Refs.~\cite{gould1988direct,gould1991gravitational} studied analytically the dynamics of diffusion throughout all Earth-orbit-crossing phase space (but without account of orbital resonances), in the context of DM capture into the Solar System (see also Refs.~\cite{lundberg2004weakly,peter2009dark}).  The dominant Solar basin injection for that part of phase space is those orbits which reach aphelion at or just above $1\,\mathrm{AU}$, and which therefore have nearly vanishing speed in the heliocentric frame when they cross Earth's orbital radius, or a velocity of $v_\oplus \sim 30\,\mathrm{km/s}$ (opposite Earth's motion) in the geocentric frame. Ref.~\cite{gould1991gravitational} found the combined action of Venus and Earth leading to a time $\tau_\mathrm{diff} \sim (M_\odot/M_\oplus)^2 P_\oplus$ (with $P_\oplus \equiv 1\,\mathrm{y}$) to diffuse out of this region, i.e.~longer than the age of the Solar System, leaving this part of phase space ``unfilled'' from Galactic DM capture (inner solid contour of Fig.~3 in Ref.~\cite{gould1991gravitational}). By the time-reversed argument, it would also not be depleted over that time. As these orbits do not cross Jupiter's, energy pumping due to the Jovian disturbing potential also does not occur before third order in secular perturbation theory away from orbital resonances~\cite{tisserand1889traite}, so at least as slow as $\tau_\mathrm{pump} \sim P_\mathrm{J}^4 / P_\oplus^3 \gtrsim 3 \times 10^9 \,\mathrm{y}$, where $P_\mathrm{J} \equiv (a_\mathrm{J}^3/G_N M_\mathrm{J})^{1/2}$ with $M_\mathrm{J}$ and $a_\mathrm{J}$ the Jovian mass and semi-major axis. Preliminary simulations~\cite{wiser_draft} confirm that orbits in Ref.~\cite{gould1991gravitational}'s ``unfilled'' region typically do remain there for $4.6 \times 10^9\,\mathrm{y}$, even given secular and resonant perturbations.

I therefore reckon that the actual gravitational ejection time is somewhere between the fiducial value of Eq.~\ref{eq:taueject2} based on forward simulations of (biased) NEO orbits and the maximally optimistic estimate
\begin{align}
\tau_\mathrm{eject}^\odot(\mathrm{AU}) \sim 4.5 \times 10^9\,\mathrm{y} \quad \text{(optimistic)}. \label{eq:taueject3}
\end{align}
Nevertheless, this work prudently uses Eq.~\ref{eq:taueject1} for the purposes of recasting limits. The widely varying benchmarks of Eqs.~\ref{eq:taueject1}, \ref{eq:taueject2}, and \ref{eq:taueject3} are clearly not satisfactory, and analytic arguments are unlikely to settle this issue. Dedicated simulations are sorely needed to establish not just the typical ejection time, but also to characterize the statistical behavior of $\rho_\mathrm{b}$ as well as its temporal intermittency and modulation due to Earth's weakly eccentric orbit.

\section{Soft Emission Near Mass Threshold}
For a general emission process of a weakly coupled boson with four-momentum $k = (\omega, \vect{k})$, the net energy loss rate per unit volume is:
\begin{align}
Q = \iiint \dd \mathbb{P}_\mathrm{in} \dd \mathbb{P}_\mathrm{out} \frac{\ddbar^3 \vect{k}}{2} \, \overline{|\mathcal{M}|^2} \deltabar^4(P_\mathrm{in} - P_\mathrm{out} - k) \mathbb{F}. \label{eq:Qgeneral}
\end{align}
where  $\overline{|\mathcal{M}|^2}$ is the spin-averaged square matrix element, $P_\mathrm{in} \equiv (E_\mathrm{in}, \vect{P}_\mathrm{in}) = \sum_i^\mathrm{in} p_i$ is the total four-momentum of incoming particles, $\int \dd \mathbb{P}_\mathrm{in} \equiv \prod_i^\mathrm{in} g_i \int \ddbar^3 \vect{p}_i/(2E_i)$ are the usual momentum integral measures (similarly, $P_\mathrm{out}$ and $\int \dd \mathbb{P}_\mathrm{out}$ for outgoing particles), with factors $g_i$ accounting for internal degrees of freedom such as spins. Shorthand $\ddbar \equiv \dd / (2\pi)$ and $\deltabar() \equiv 2\pi \delta()$ is used. Generalization to multi-particle emission, e.g.~pair production of neutrinos, is straightforward. For some models of weakly coupled particles (such as kinetically mixed photons), special care must be taken to include medium-dependent dispersion relations and interactions, complications ignored here.

The phase-space distribution functions---$f(\vect{k})$ of the weakly coupled particle and $f_i(\vect{p}_i)$ of the in- and out-going particles---are collected into the function:
\begin{align}
\hspace{-0em} \mathbb{F} =  (1+f) \mathbb{F}_\mathrm{in} \overline{\mathbb{F}}_\mathrm{out} - f \mathbb{F}_\mathrm{out}\overline{\mathbb{F}}_\mathrm{in} \equiv \mathbb{F}_\mathrm{sp} + \mathbb{F}_\mathrm{st} - \mathbb{F}_\mathrm{abs}.
\end{align}
The three terms correspond to spontaneous emission, stimulated emission, and absorption, respectively. Lastly, I define $\mathbb{F}_\mathrm{in} \equiv \prod_i^\mathrm{in} f_i$ and $\overline{\mathbb{F}}_\mathrm{in} \equiv \prod_i^\mathrm{in} (1 + \eta_i f_i)$, with $\eta_i = \pm 1$ if particle $i$ is a boson/fermion to include Bose enhancement or Pauli blocking. All distribution functions are normalized such that $\langle n_i \rangle = g_i \int \ddbar^3 \vect{p}_i f_i(\vect{p}_i)$ gives the local number density. Those of the in- and outgoing particles are assumed to be thermally equilibrated at the same temperature $T$:
\begin{align}
f_i(\vect{p}_i) = \frac{1}{e^{(E_i - \mu_i)/T} - \eta_i}, \label{eq:distros}
\end{align}
with (relativistic) chemical potentials $\mu_i$.

At early times, when the stellar basin is nearly empty $f \ll 1$, only spontaneous emission is operable---the limit taken in the case study of the next section. At non-negligible occupation numbers $f \gtrsim 1$, stimulated emission and absorption should be considered---in equilibrium (i.e.~using Eq.~\ref{eq:distros}), they famously cancel each other in the limit of zero recoil~\cite{1916DPhyG..18..318E,1917PhyZ...18..121E}. However, at finite $\omega$, absorption always ``wins'' eventually:
\begin{align}
\frac{\mathbb{F}_\mathrm{st} - \mathbb{F}_\mathrm{abs}}{f} &=  \prod_i^\mathrm{in} f_i \prod_j^\mathrm{out} (1 + \eta_j f_j) -  \prod_j^\mathrm{out} f_j \prod_i^\mathrm{in} (1 + \eta_i f_i) \nonumber \\
&=\mathbb{F}_\mathrm{sp} \left(1 - e^{\omega/T} \right) < 0,
\end{align}
where Eq.~\ref{eq:distros} and conservation of both energy and chemical potential were used to get to the second line. Therefore, any (bound) mode $\vect{k}$ eventually saturates to a level $f(\vect{k}) =1/(e^{\omega/T} - 1)$ where $\mathbb{F} = 0$ and detailed balance is reached. (Of course, unbound modes with $|\vect{k}|/m > v_\mathrm{esc}$ continue to escape to infinity and never reach this level.) The saturation condition implicitly defines the absorption timescale
\begin{align}
\tau_\mathrm{abs} \sim \frac{1}{\dot{\rho}_\mathrm{b}} \frac{m^4 v_\mathrm{esc}^3}{e^{m/T} - 1}, \label{eq:tauabsorb}
\end{align}
roughly the time it takes for the stellar core to reach densities of order the second fraction. The effective absorption time for locations far outside the star is parametrically longer for stars with planets, which will secularly perturb the emitted particles' trajectors into non-star-traversing orbits, and thus diffuse into other parts of phase space with lower occupation numbers. I treated the case for single boson emission; occupation levels for fermions would saturate earlier, at $f = 1/(e^{\omega/T}+1)$. I am not aware of a regime in \emph{dense} astrophysical media where Bose-enhanced exponential growth can occur (like astrophysical masers in dilute molecular gas clouds), as $\mathbb{F}_\mathrm{st} > \mathbb{F}_\mathrm{abs}$ necessitates population inversion, and thus large departures from thermal equilibrium. 

Finally, processes with constant matrix elements in the soft limit, i.e.~$\overline{|\mathcal{M}|^2} \simeq \mathrm{const.}$ as $\vect{k} \to 0$, are precisely those with $ n_p = 1$ that efficiently fill stellar basins. In the soft limit, one can ignore $\vect{k}$ in the three-momentum delta function and replace $\omega$ with $m$ everywhere in Eq.~\ref{eq:Qgeneral}. This allows the phase space integral to be factored out $\int \ddbar^3 \vect{k} \simeq (m^3/\sqrt{2}\pi^2) \int \dd \tilde{\omega}_k\, \tilde{\omega}_k^{1/2}$, realizing the form of Eq.~\ref{eq:tildeQ} with $n_p = 1$.

\section{Solar Axion Basin}
As a case study of interest, consider a pseudoscalar particle $a$ (hereafter ``axion'') coupled derivatively to the axial electron current with dimensionless coupling $g_{aee}$:
\begin{align}
\mathcal{L} = \frac{(\partial a)^2}{2} - \frac{m^2 a^2}{2}+ \frac{g_{aee}}{2 m_e} (\partial_\mu a) \bar{\psi}_e \gamma^\mu \gamma^5 \psi_e. \label{eq:lagrangian}
\end{align}
If the axion mass $m$ is below twice the electron mass $m_e$, its main radiative decay channel is to two photons (minimally induced from a diagram with an electron loop and the  $a\bar{\psi}_e \psi_e$ vertex from Eq.~\ref{eq:lagrangian}), with a radiative rate
\begin{align}
\tau_\mathrm{rad}^{-1} \simeq \frac{\alpha^2 g_{aee}^2 }{9216 \pi^3} \frac{m^7}{m_e^6} &\approx \frac{1}{2 \times 10^{26}\,\mathrm{y}} \left[\frac{g_{aee}}{10^{-13}}\right]^{2}\left[\frac{m}{\mathrm{keV}}\right]^{3}, \label{eq:taurad}
\end{align}
assuming $m \ll 2 m_e$~\cite{nakayama2014anomaly}. The numerical estimate is evaluated near parameters of interest, at which the axion can be considered cosmologically stable, and the X-ray photon rate is beyond current observational capabilities~\cite{Pospelov_2008,takahashi2020xenon1t}. The decay to two photons could be even further suppressed (at the expense of tuning), but could also be much higher if the Peccei-Quinn symmetry is anomalous and the axion has a direct coupling to photons~\cite{Pospelov_2008}. Observable implications of rapid decays of stellar basin axions to photons are left to future work.

In the remainder of this section, I compute the contributions to the local basin density $\rho_\mathrm{b}$ from bremsstrahlung in the Sun and Earth's core, which are the primary sources for basin axions with masses below $5\,\mathrm{keV}$. Photoproduction (the Compton process) in the Sun is the main production channel at higher masses. (I have not yet included other atomic processes, such as those of Ref.~\cite{redondo2013Solar}, within this framework.) I first calculate $\tilde{Q}_p$ for each process (massage Eq.~\ref{eq:Qgeneral} into the form of Eq.~\ref{eq:tildeQ}), and subsequently evaluate the volume integrals of Eq.~\ref{eq:rhodot} over the Standard Solar Model of Ref.~\cite{bahcall2005new} and the Preliminary Reference Earth Model of Ref.~\cite{dziewonski1981preliminary}. 

\begin{figure*}
\includegraphics[width = 0.8 \textwidth]{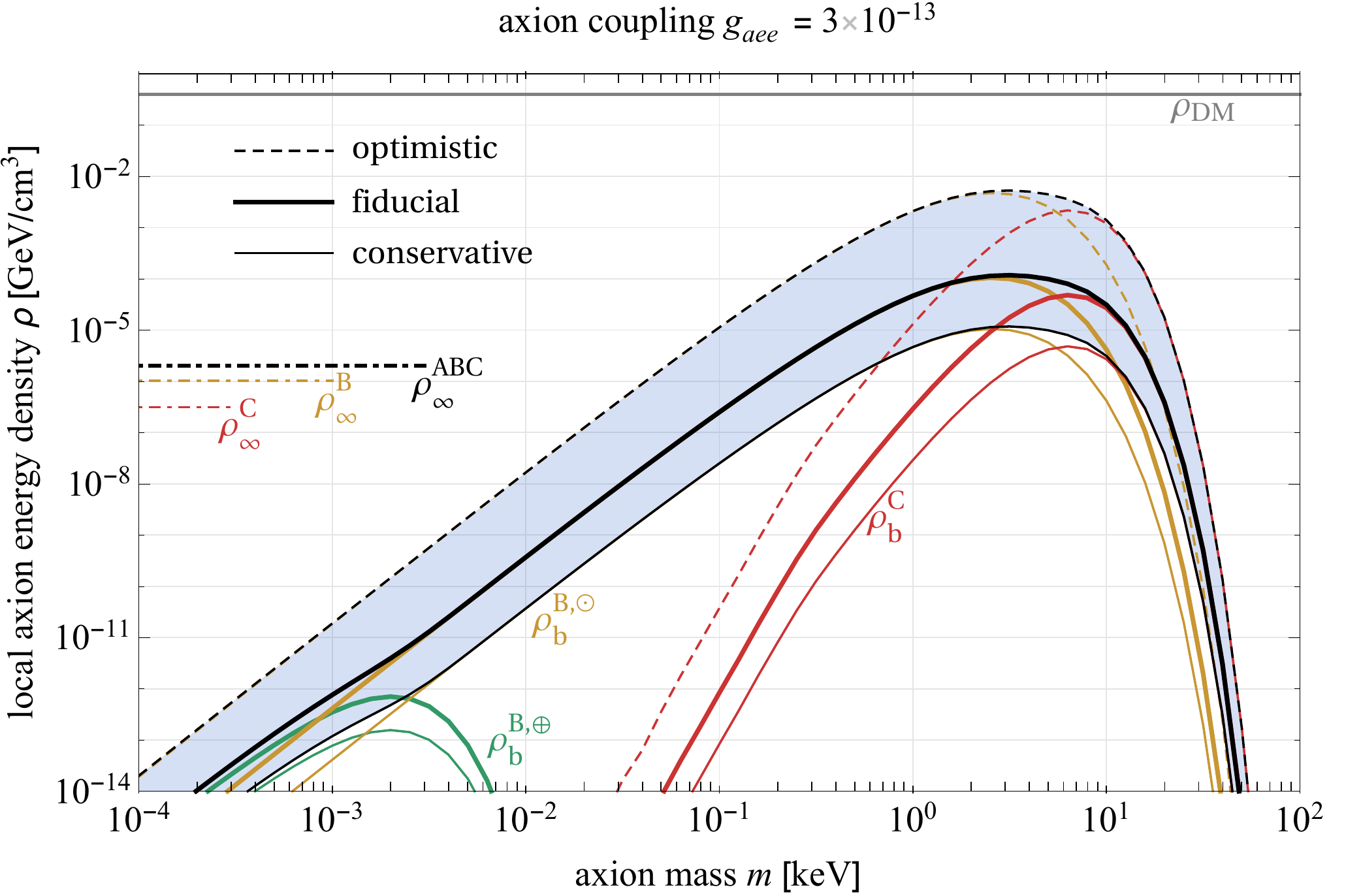}
\caption{Axion energy density at Earth's surface from the Solar basin ($\rho_\mathrm{b}$) and from the relativistic Solar flux ($\rho_\infty$) as a function of axion mass $m$, for a reference axion-electron coupling of $g_{aee} = 3 \times 10^{-13}$. Solar basin contributions from the Compton process (red, $\rho_\mathrm{b}^\mathrm{C}$) and bremsstrahlung (blue, $\rho_\mathrm{b}^{\mathrm{B},\odot}$) are shown for three benchmark basin lifetimes (Eqs.~\ref{eq:taueject1}, \ref{eq:taueject2}, \ref{eq:taueject3}; respectively thin, thick, dashed). Black curves are totals of these subprocesses, including a component $\rho_\mathrm{b}^{\mathrm{B},\oplus}$ at low energies from bremsstrahlung in the Earth's core (for $\tau = 1,5 \times 10^9\, \mathrm{y}$). The blue shaded region spans the range of these total predictions; its conservative lower boundary is used to set conservative upper limits on couplings in Fig.~\ref{fig:gbasin}. Relativistic Solar axion flux energy densities (in the massless limit, no attempt was made to compute finite-mass suppression) are shown as dot-dashed lines. The Compton ($\rho_\infty^\mathrm{C}$) and bremsstrahlung ($\rho_\infty^\mathrm{B}$) values agree with those of Ref.~\cite{redondo2013Solar}, whose luminosity ratios are used to get $\rho_\infty^\mathrm{ABC}$ that also includes atomic recombination subprocesses. For axion masses $m \gtrsim 0.5\,\mathrm{keV}$ (and perhaps as low as tens of eV), the local Solar basin energy density exceeds that of the relativistic Solar flux, and may reach up to 1\% of the cosmic DM energy density $\rho_\mathrm{DM}$ (gray line) at this reference coupling.} \label{fig:rhobasin}
\end{figure*}

The results in terms of energy density at Earth's surface are displayed in Fig.~\ref{fig:rhobasin} for a reference coupling of $g_{aee} = 3 \times 10^{-13}$. Evidently, the (coupling-independent) parametric estimate of $\rho_\mathrm{b}/\rho_\infty$ in Eq.~\ref{eq:rhoparametric} holds, validating the conclusion that the local Solar axion density is dominated by the Solar basin for keV-scale axion masses, at least by one order (possibly four orders) of magnitude. The Solar basin densities remain low enough that one can ignore axion re-absorption at Earth's location (cfr.~Eq.~\ref{eq:tauabsorb}). The reader not interested in the details of the computation may skip to the next section, where it is shown that the basin energy densities of Fig.~\ref{fig:rhobasin} are potentially detectable in current direct detection experiments.

\subsection{Bremsstrahlung}

For axion bremsstrahlung production $e^- + Z_j \rightarrow e^- + Z_j + a$ from collisions of electrons with nuclear ions of charges $Z_j$ and masses $m_j$, and corresponding momenta $p_1 + p_2 = p_3 + p_4 + k$, the squared matrix element in the soft limit is:
\begin{align}
\overline{|\mathcal{M}_{\mathrm{B},j}|^2} = \frac{Z_j^2 e^4 g_{aee}^2}{\vect{q}^2 + \kappa_s^2} \frac{4 m_j^2}{m_e^2} .
\end{align}
Screening effects are accounted for by the Debye-H\"uckel scale $\kappa_s^2 \simeq (4 \pi \alpha / T)(\sum_j n_j Z_j^2 + n_e) \equiv 4 \pi \alpha (\overline{n}_N + n_e) / T $ , proportional to the number density of different ions ($n_j$) and electrons ($n_e$). The three-momentum transfer is defined as  $\vect{q} \equiv \vect{p}_2 - \vect{p}_4$. The ions have low occupation numbers $f_2, f_4 \ll 1$, and may be considered infinitely heavy  relative to electrons so that nuclear recoil can be neglected. That leaves a trivial $\vect{p}_2$ integral, $\vect{p}_3 = \vect{p}_1 - \vect{q}$ fixed by the delta function (ignoring axion momenta in the soft limit $\vect{k} \to 0$, $\omega \to m$), and the integral over $\vect{p}_4$ can be replaced by one over $\vect{q}$. Plugging everything into Eq.~\ref{eq:Qgeneral} yields:
\begin{align}
\hspace{-0em}&\frac{\dd Q_\mathrm{B}}{\ddbar^3 \vect{k}} = \frac{\overline{n}_N e^4 g_{aee}^2}{8 m_e^4} \hs \iint \hs \ddbar^3 \vect{p}_1 \ddbar^3 \vect{q} \frac{\deltabar\left[ \frac{\vect{p}_1^2 - \vect{p}_3^2}{2m_e} - m \right]}{\vect{q}^2 + \kappa_s^2} 2 f_1 (1-f_3) \nonumber \\
&= \frac{\overline{n}_N e^4 g_{aee}^2}{32 \pi^3 m_e^2}  \int_{m}^\infty \hs \hs \dd \omega_p \,  f_1 (1-f_3) \ln\frac{2 + 2 \sqrt{1-\epsilon} - \epsilon + \xi}{\epsilon + \xi}  . \label{eq:Qbrem}
\end{align}
In the first line, the electrons are taken to be nonrelativistic. The second line makes use of the following definitions: $\omega_p \equiv \vect{p}_1^2 / 2 m_e$ as the kinetic energy of the incoming electron, $\epsilon \equiv m/\omega_p$ as the ratio of axion mass over available electron kinetic energy, and $\xi \equiv \kappa_s^2 / 2 m_e \omega_p$ as a dimensionless screening measure. 

\paragraph*{Nondegenerate Medium.---}
Electrons in the Solar plasma are nondegenerate, as $n_e \ll (m_e T)^{3/2}$, so one can approximate $f_3 \simeq 0$ (no Pauli blocking) and Eq.~\ref{eq:distros} as a Maxwell-Boltzmann distribution:
\begin{align}
2 f_1(\vect{p}_1) \simeq n_e \left( \frac{2 \pi}{m_e T}\right)^{3/2} \exp\left\lbrace - \frac{\omega_p}{T}\right\rbrace.
\end{align}
With these approximations and after converting $\int \ddbar^3 \vect{k} \simeq (m^3/\sqrt{2}\pi^2) \int \dd \tilde{\omega}_k\, \tilde{\omega}_k^{1/2}$  as at the end of the previous section, comparison of Eq.~\ref{eq:Qbrem} with Eq.~\ref{eq:Qgeneral} yields:
\begin{align}
\hspace{-0.5em}\tilde{Q}^\mathrm{ND}_\mathrm{B} \simeq \frac{\alpha^2 g_{aee}^2}{2\pi^{3/2}} \frac{\overline{n}_N n_e  m^3}{m_e^{7/2} T^{1/2}}  \int_0^1 \dd \epsilon \,  \frac{\ln\frac{2 + 2 \sqrt{1-\epsilon} - \epsilon + \xi}{\epsilon + \xi}  }{\exp\lbrace \frac{m}{\epsilon T} \rbrace}. \label{eq:QbremND}
\end{align}
The screening measure $\xi$ is quite small in practice. I find that the integral in Eq.~\ref{eq:QbremND} is approximated by the empirical formula $4.7 \exp\lbrace -1.38 s^{1/2} - 0.704 s - 0.024 s^{3/2} \rbrace$ for $s \equiv m/T \lesssim 20$ to better than $\pm 5\%$ accuracy throughout the most strongly emitting regions of the Solar interior.
Axion production from electron-electron bremsstrahlung gives another additive contribution, with the same result provided one make the replacement $\overline{n}_N \leftrightarrow n_e/\sqrt{2}$, similar as in Ref.~\cite{raffelt1986astrophysical}. The results from the volume integral of Eq.~\ref{eq:rhodotsimple} with substitution of Eq.~\ref{eq:QbremND} over the Solar Model of Ref.~\cite{bahcall2005new} are plotted in Fig.~\ref{fig:rhobasin} as the gold curves. 

For reference, the total energy loss per unit volume in the massless axion limit is
\begin{align}
Q = \frac{32\sqrt{2} \alpha^2 g_{aee}^2}{45 \pi^{3/2}} \frac{n_e (\bar{n}_N + n_e/\sqrt{2}) T^{5/2}}{m_e^{7/2}},
\end{align}
with screening corrections ignored~\cite{raffelt1996stars}. The luminosity found by integrating over the Solar volume leads to a relativistic flux energy density $\rho_\infty^\mathrm{B}$ plotted as the gold dashed curve in Fig.~\ref{fig:rhobasin} (cfr.~Eq.~\ref{eq:rhoinf} and integral above).

\paragraph*{Degenerate Medium.---}
Earth's core also sources a local ``planetary basin" of axions via electron-ion bremsstrahlung. It consists primarily of metallic iron, wherein the electron gas is strongly degenerate, with a Fermi energy $E_F = p_F^2/2 m_e = \mu - m_e \gg T$. The electron density is $n_e = p_F^3/(3 \pi^2)$. At standard conditions, iron's mass density is $\rho_N  = m_N n_N = 7.874\,\mathrm{g}\,\mathrm{cm}^{-3}$, with Fermi energy  $E_F \approx 11.1\,\mathrm{eV}$, Fermi momentum $p_F \approx 3.37\,\mathrm{keV}$, and electron density $n_e \approx 0.411 \, \mathrm{keV}^{-3}$. From charge neutrality, I derive an effective ionic charge $Z_N \simeq n_e / n_N \approx 0.636$ (the core electrons of iron shield the rest of the nuclear charge). I adopt a Thomas-Fermi approximation for the screening scale: $\kappa_s \simeq (e/\pi) \sqrt{m_e  p_F} \approx 4.00\,\mathrm{keV}$. One complication is that Earth's core is considerably denser than iron at ambient pressures; I deal with this by keeping $Z_N$ fixed and scaling up the density and Fermi levels by appropriate factors. I ignore nickel and other elements, but they would be straightforward to include by summing over $N = \mathrm{Fe}, \mathrm{Ni}, \dots$. I take a fiducial temperature for both the inner and outer core of $T_\oplus \approx 5000\,\mathrm{K}$, and the mass densities from Ref.~\cite{dziewonski1981preliminary}.

For a strongly degenerate medium, the result of Eq.~\ref{eq:Qbrem} simplifies considerably because $\epsilon \simeq m / E_F$ may be taken to be much less than unity, as high $\epsilon$ values would be Boltzmann suppressed. I find the closed-form expression:
\begin{align}
\tilde{Q}^\mathrm{D}_\mathrm{B} \simeq \frac{\alpha^2 g_{aee}^2}{2^{3/2} \pi^3} \frac{Z_N^2 n_N}{m_e^2} \frac{m^4}{e^{m/T}-1}\ln\left[ 1 + \frac{\pi p_F}{\alpha m_e} \right].
\end{align}
Electron-electron bremsstrahlung production of axions can be ignored entirely, as it is suppressed by the Pauli blocking factor $\int \ddbar^3 \vect{p}_2 \, 2 f_2 (1-f_4) / n_e \simeq 3 E_F T / p_F^2 \ll 1$. The associated basin energy density at $R = R_\oplus \approx 6371\,\mathrm{km}$ is plotted as green curves in Fig.~\ref{fig:rhobasin}.

\subsection{Compton Process}
At high axion masses significantly above the temperature (and plasma frequency) of the Sun, the Compton process $\gamma + e^- \rightarrow a + e^-$ is the dominant axion production channel, with an energy production per unit of 6D phase space volume of:
\begin{align}
\hspace{-0em}\frac{\dd Q_\mathrm{C}}{\ddbar^3 \vect{k}} = 8 \Bigg[ \prod_{i}^{\gamma, 1,2} \int \frac{\ddbar^3 \vect{p}_i}{2E_i} \Bigg] \overline{|\mathcal{M}_{\mathrm{C}}|^2} \deltabar^4( P)f_\gamma f_1 (1-f_2)
\end{align}
with the abbreviation $P \equiv p_\gamma + p_1 - k - p_2$. The polarization-averaged square matrix element is $\overline{|\mathcal{M}_\mathrm{C}|^2|} \simeq e^2 g_{aee}^2 m^2 / m_e^2$ in the soft limit. The electrons are taken to be nonrelativistic ($E_{1,2} \simeq m_e$), nondegenerate ($f_2 \simeq 0$) bystanders that absorb momentum but have negligible energy recoil. Then the $\vect{p}_1$ integral over $2 f_1$ gives a factor of $n_e$, the $\vect{p}_2$ integral is trivial, and the $\vect{p}_\gamma$ integral over $f_\gamma$ and $\delta(E_\gamma - m)$ can be done analytically, taking into account the dispersion relation $E_\gamma^2 = \omega_\mathrm{pl}^2 + p_\gamma^2$ at plasma frequency $\omega_\mathrm{pl}^2 = 4 \pi \alpha n_e / m_e$. I find:
\begin{align}
\tilde{Q}_\mathrm{C} \simeq \frac{\alpha g_{aee}^2}{2^{3/2} \pi^2} \frac{n_e}{m_e^4}\frac{ m^5 \sqrt{m^2 - \omega_\mathrm{pl}^2}}{e^{m/T}-1},
\end{align}
kinematically allowed only when $m > \omega_\mathrm{pl}$, which is approximately $0.3 \, \mathrm{keV}$ in the Solar core. The resulting Solar basin energy densities from this process are depicted in Fig.~\ref{fig:rhobasin} by red curves.

For reference, the total energy loss rate per unit volume from the Compton process is
\begin{align}
Q_\mathrm{C} = \frac{40 \zeta(6) \alpha g_{aee}^2}{\pi^3} \frac{n_e T^6}{m_e^4}
\end{align}
as calculated for massless axions in Ref.~\cite{raffelt1986astrophysical}. Its Solar volume integral leads to a relativistic flux energy density shown as the red dashed curve in Fig.~\ref{fig:rhobasin}.

\section{Direct Detection Signals}

Given that energy densities of the Solar axion basin at Earth's location can be competitive or exceed those of the relativistic (unbound) Solar axion flux, searches for Solar axions must be recast to take into account this ``new'' population component. To first order, the stellar basin resembles a DM halo with a peculiar density profile centered on the star that is continually replenished. At Earth's location within the Solar System, the velocity dispersion within the basin would be significantly smaller---bounded by $[v^\odot_\mathrm{esc}(\mathrm{AU})]^2 \sim 10^{-8}$ as opposed to $[v_\mathrm{DM}^\mathrm{galactic}]^2 \sim 10^{-6}$. For all practical purposes, any one axion absorption event from the Solar basin would thus be indistinguishable from an axion DM absorption event.

\begin{figure*}
\includegraphics[width = 0.95 \textwidth]{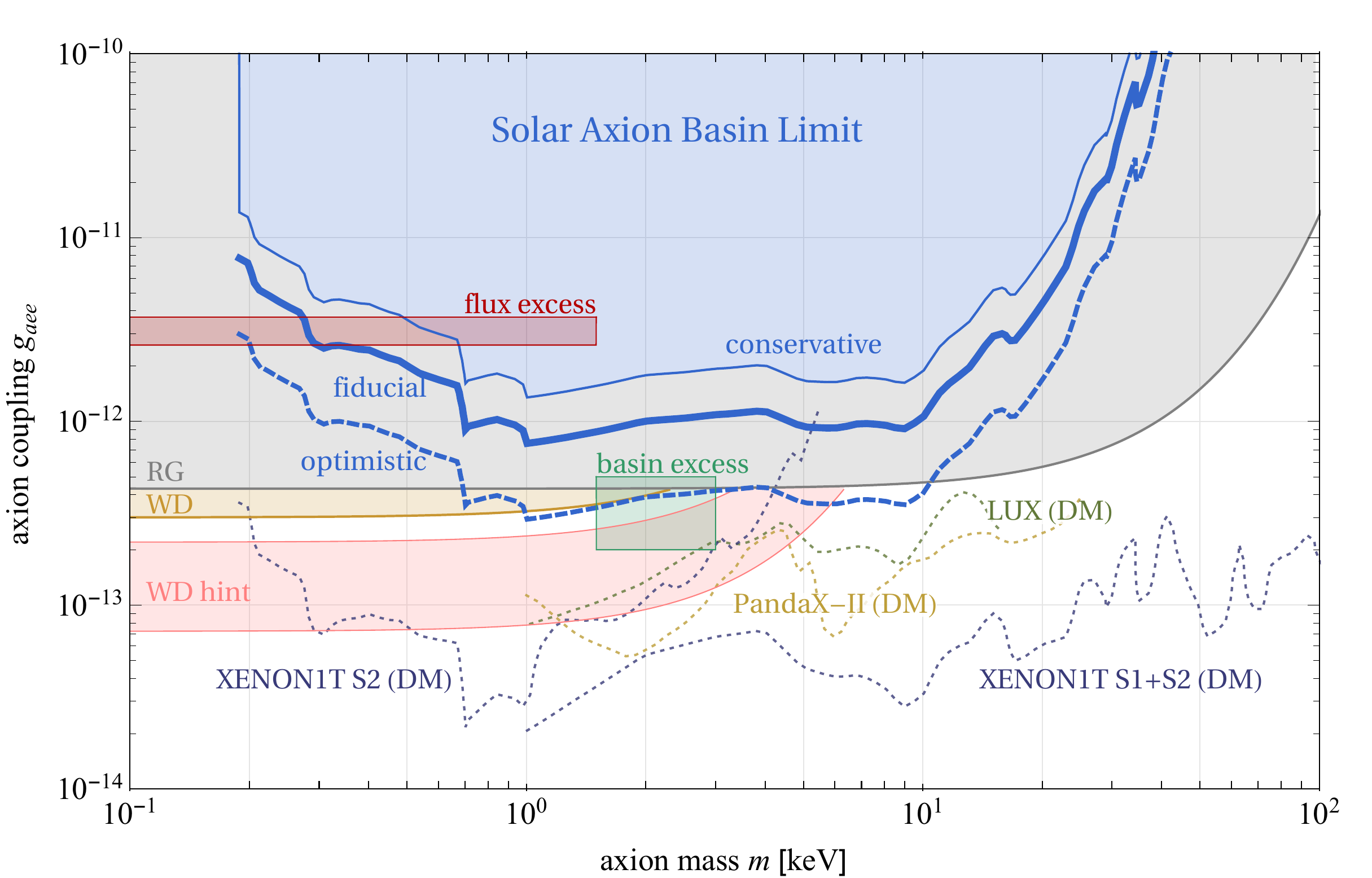}
\caption{Limits on the pseudoscalar-electron coupling $g_{aee}$ of axions as a function of mass $m$. The blue region is conclusively excluded by the recasting procedure of Eq.~\ref{eq:recast} with the conservative estimate of the gravitational ejection time of Eq.~\ref{eq:taueject1}, independent of assumptions about dark matter. This limit is a new ``Solar basin'' interpretation of dark-matter axion absorption line searches by  the XENON1T~\cite{aprile2019light,aprile2020observation}, LUX~\cite{akerib2017first}, and PandaX-II~\cite{fu2017limits} collaborations (upper limits, assuming the axion makes up all DM, depicted as thin dotted lines). The region above the thick blue curve is also disfavored if the fiducial estimate of Eq.~\ref{eq:taueject2} is adopted. Both XENON1T's S2-only~\cite{aprile2019light} and S1+S2~\cite{aprile2020observation} searches could have plausibly detected an axion in the Solar basin near or above the dashed blue curve (with the optimistic estimate of Eq.~\ref{eq:taueject3}), and below anomalous cooling constraints of white dwarfs (WD) and red giants (RG) (orange and gray regions) extrapolated to high axion masses. The tentative excess of Ref.~\cite{aprile2020observation} is in gross violation of those cooling constraints when attributed to absorption of a relativistic Solar axion flux (red region), but could be mutually compatible with absorption of Solar basin axions in the $m \in [1.5,3] \, \mathrm{keV}$ range (inside the green rectangle), as well as with hints of anomalous WD cooling (pink region).} \label{fig:gbasin}
\end{figure*}

Dark-matter absorption line searches such as those of Refs.~\cite{aprile2019light,aprile2020observation} that set limits $g_{aee}^{\mathrm{DM}}$ on the electron-axion coupling (depicted as thin lines in Fig.~\ref{fig:gbasin}) may thus be recast as coupling limits $g_{aee}^{\mathrm{basin}}$ that do \emph{not} make use of the assumption that the axions make up the cosmic DM. Suppose the DM event rate is $ C (g_{aee}^{\mathrm{DM}})^2 \rho_\mathrm{DM}$ for some proportionality constant $C$. Then the event rate for Solar basin absorption is $C (g^\mathrm{basin}_{aee})^4\dot{\rho}_\mathrm{b}\big|_{g=1} \tau$ with the same constant $C$. Equating the two produces the recasting map:
\begin{align}
\left|g^\mathrm{basin}_{aee}\right| \cong \left|g^\mathrm{DM}_{aee}\right|^{1/2} \left(\frac{ \rho_\mathrm{DM}}{\dot{\rho}_\mathrm{b}\big|_{g=1} \tau }\right)^{1/4}. \label{eq:recast}
\end{align}
Above, $\dot{\rho}_\mathrm{b}\big|_{g=1}$ is the energy injection rate at $g_{aee} = 1$ (merely a mathematical trick: at such high couplings the Sun would become opaque to axions). The resulting Solar axion basin limit in Fig.~\ref{fig:gbasin} is depicted by the light blue region (for the conservative estimate of $\tau$ in Eq.~\ref{eq:taueject1}); the lower blue lines should be viewed as delineating parameter space that is disfavored (``fiducial'', with Eq.~\ref{eq:taueject2}) and where signals are conceivable (``optimistic'', with Eq.~\ref{eq:taueject3}).

Fig.~\ref{fig:gbasin} displays cooling constraints from WDs~\cite{Bertolami:2014wua} and RGs~\cite{viaux2013neutrino} as orange and gray regions respectively. They are crudely extrapolated to finite axion masses by a factor proportional to the inverse square root of $\int_m^\infty \dd \omega \, \omega^2 \sqrt{\omega^2 - m^2}/(e^{\omega/T}-1)$ as in Refs.~\cite{raffelt1990astrophysical,calibbi2020looking}.
This extrapolation might be overly stringent, as the temperature variation among white dwarfs in particular will skew cooling analyses based on luminosity functions. Colder WDs are more strongly affected by the finite axion mass than hotter ones, and because they are more numerous, these are also the better constrained observationally. There may be some indications of excess WD cooling, both from luminosity distribution~\cite{Isern_2008,Bertolami:2014wua} and pulsation period drift~\cite{isern2010axions,C_rsico_2012,C_rsico_2012b} analyses, that could potentially be explained by an electron-axion coupling; this parameter space is plotted as the pink region below the constraints. On the other hand, Ref.~\cite{hansen2015constraining} has contested these cooling hints, leveraging a potentially more sensitive population of hot WDs in a single globular cluster, and claims a limit of $g_{aee} \lesssim 8.3 \times 10^{-14}$, just above the lower boundary of the pink region in Fig.~\ref{fig:gbasin}.  

The low-energy excess of electron-recoil events recently reported by the XENON1T experiment~\cite{aprile2020observation} is not compatible with stellar cooling constraints when interpreted as a signal from the relativistic Solar axion flux---sketched as the red ``flux excess'' region in Fig.~\ref{fig:gbasin} for $m \lesssim 1.5 \, \mathrm{keV}$ and $2.6 \times 10^{-12} \lesssim g_{aee} \lesssim 3.7 \times 10^{-12}$. Note that the conservative Solar axion basin limit (recast from the S2-only DM analysis from the same collaboration) further excludes $m \gtrsim 0.7 \, \mathrm{keV}$ within this space. If the optimistic estimate for the gravitational ejection time scale in Eq.~\ref{eq:taueject3} is indeed correct, then the excess of events could instead be due to an absorption line from the Solar basin if the axion mass were roughly between $1.5$--$3\,\mathrm{keV}$, a region highlighted as the ``basin excess'' green box in Fig.~\ref{fig:gbasin}. The global significance for the axion DM line search was reported~\cite{aprile2020observation} to be lower primarily due to the look-elsewhere-effect within the search region $m \in [1,210]\,\mathrm{keV}$; Fig.~\ref{fig:gbasin} shows that the region of interest for a Solar basin axion search should instead be $m \lesssim 20\,\mathrm{keV}$, with a correspondingly lower trials factor and higher global significance. With a generous stretch of the imagination, these axion parameters might also be responsible for the WD cooling hint, though the analysis of Ref.~\cite{hansen2015constraining} disfavors this scenario. 

I conclude this section with the main takeaway: axio-electric direct detection experiments are much more sensitive to keV-scale Solar axions than previously anticipated. If simulations confirm the optimistic estimate for the gravitational ejection time of Eq.~\ref{eq:taueject3}, then Fig.~\ref{fig:gbasin} shows that the line search of Ref.~\cite{aprile2020observation} already has the best discovery potential to $g_{aee}$ for $m \in [6,10]\,\mathrm{keV}$, surpassing the strongest claimed WD and RG cooling bounds~\cite{hansen2015constraining,Bertolami:2014wua, viaux2013neutrino}.

\section{Discussion}

I have identified and described a previously overlooked process in astroparticle physics. Amusingly, it is not just feeble interaction strengths, which allow for volumetric stellar emission, that make stars interesting sources of weakly coupled particles. A small but non-zero mass opens up the possibility of bound-orbit emission that fills up a ``stellar basin'' over time, a spectacular channel not available to massless photons. I have performed a case study for CP-odd pseudoscalars coupled to electrons in some detail, and investigated the consequences for direct detection experiments.

It would be interesting to study stellar basins in the context of other particles, couplings, and experiments, as well as for stars other than the Sun. Other direct detection experiments---such as CAST~\cite{cast2017}, which probes the axion-photon coupling---might also be affected by the Solar basin. Preliminary estimates indicate that axions  would quickly saturate to maximum occupation numbers (see Eq.~\ref{eq:tauabsorb}) around isolated neutron stars from nuclear bremsstrahlung production, even with incredibly tiny couplings to nuclear matter. This ``neutron star axion basin'' could potentially boost indirect detection signatures such as those of e.g.~Ref.~\cite{buschmann2019x,dessert2019x}. While the effects on initial cooling rates are negligible in practice, the basin could potentially act as a reservoir that \emph{slows} the cooling process, as the basin is re-absorbed by the star when it cools down. If part of the basin is protected on a long-lived orbit outside the stellar interior, secular perturbations may kick it back inside (or into other surrounding objects) at a later time. The Sun also fills up a ``neutrino basin'', but at energy densities far below that of even the cosmic neutrino background, rendering detection all but impossible. 

Finally, the most pressing loose end in the present analysis is the poorly known gravitational ejection time scale. Simulations of the orbital dynamics (beyond several Lyapunov times) are needed to characterize the expected Solar basin density and its fluctuations, which would be crucial input to direct detection experiments. Solar basin signals (with $\rho_\mathrm{b} \propto 1/R^4$) will exhibit stronger annual modulation than Solar flux signals (with $\rho_\infty \propto 1/R^2$) due to Earth's eccentric orbit, which is also out of phase with the modulation expected from a DM signal. This temporal variation combined with the ultra-narrow kinetic energy spread would be a smoking gun of direct detection signals from the Solar basin.

\medskip

\acknowledgments{KVT is especially indebted to Timothy Wiser for sharing insights regarding gravitational ejection dynamics and phase space diffusion, and thanks Asimina Arvanitaki, Masha Baryakhtar, Nathaniel Craig, Isabel Garcia Garcia, Junwu Huang, Andrew Jayich, Ania Jayich, Amalia Madden, Cristina Mondino, Aaron Pierce, Oren Slone, Anna-Maria Taki, and Neal Weiner for discussions and comments. This research is funded by the Gordon and Betty Moore Foundation
through Grant GBMF7392, and supported in part by the National Science Foundation under Grant No.~NSF PHY-1748958.}

\bibliography{CoolBasin}

\end{document}